\documentclass[sigconf,natbib=true,9pt]{acmart}
\AtBeginDocument{%
  }

\setcopyright{acmlicensed}

\usepackage{amsmath,amssymb,amsfonts}
\usepackage{algorithmic}
\usepackage{graphicx}
\usepackage{textcomp}
\usepackage{amsmath}
\usepackage{tikz}
\usepackage{colortbl}
\usepackage{multirow}
 \usepackage[inline]{enumitem}
\usepackage{caption}
\usepackage{subcaption}
\usepackage{balance}
\definecolor{mycolorhigh}{HTML}{2D7F9D}
\definecolor{mycolorlow}{HTML}{d7ecf4}

\newcommand{\mul}[1]{%
\definecolor{mycolor}{HTML}{#1}%
\begin{tikzpicture}%
    \fill[mycolor](0,0) circle (3pt);
\end{tikzpicture}%
}

\newcommand{\precise}[1]{%
\definecolor{mycolor}{HTML}{#1}%
\begin{tikzpicture}%
     \draw[draw=mycolor, fill=mycolor] 
        (0, 0) -- 
        (0.2, 0) -- 
        (0.1, 0.2) -- 
        cycle; 
\end{tikzpicture}%
}

\newcommand{\drawsquare}[1]{%
\definecolor{mycolor}{HTML}{#1}%
\begin{tikzpicture}%
    \draw[mycolor, fill=none] (0, 0) rectangle (0.2, 0.2);Filled green dot with a radius of 3pt
\end{tikzpicture}%
}
\newcommand{\drawdot}[1]{%
\definecolor{mycolor}{HTML}{#1}%
\begin{tikzpicture}%
    \draw[mycolor] (0, 0) circle[radius=0.1cm];
\end{tikzpicture}%
}
\newcommand{\drawdiamond}[1]{%
\definecolor{mycolor}{HTML}{#1}%
\begin{tikzpicture}%
    \begin{tikzpicture}
    \draw[mycolor, fill=none] 
        (0.1, 0.2) -- 
        (0.2, 0.1) -- 
        (0.1, 0) -- 
        (0, 0.1) -- 
        cycle; 
\end{tikzpicture}
\end{tikzpicture}%
}

\newcommand{\barmulgraphic}{%
MAE Reduction
0\% \includegraphics[]{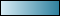} 100\%
}

\newcommand{\barexpinv}{%
1 \includegraphics[]{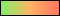} 0

}

\newcommand{\barexp}{%
0\% \includegraphics[]{results/colored-bar.pdf} 100\%

}

\newcommand{\barexpsym}{%
Close to original \includegraphics[]{results/colored-bar.pdf} Far from original

}

\usepackage{fancyhdr}

\fancypagestyle{firstpage}{%
\fancyhf{}

\lhead{This work has been accepted at the IEEE/ACM Asia and South Pacific Design Automation Conference
(ASP-DAC) 2027}
 }

\copyrightyear{2027}
\acmYear{2027}
\setcopyright{cc}
\setcctype{by}
\acmConference[ASPDAC '27]{32nd Asia and South Pacific Design Automation Conference}{January 25--28, 2027}{Tokyo, Japan}
\acmBooktitle{32nd Asia and South Pacific Design Automation Conference (ASPDAC '27), January 25--28, 2027, Tokyo, Japan}
\acmDOI{10.1145/3840404.3848340}
\acmISBN{979-8-4007-2923-2/2027/01}

\begin{document}

\title{SWAPPER: Dynamic Operand \textbf{Sw}apping in Non-commutative \textbf{App}roximate Circuits for Online \textbf{E}rror \textbf{R}eduction}

\author{Marcello Traiola$^1$, Nazar Misyats$^2$, Silviu-Ioan Filip$^1$, Remi Garcia$^1$, Angeliki Kritikakou$^1$}
\affiliation{
$^1$Univ Rennes, CNRS, Inria, IRISA, Rennes, France -- $^2$University of British Columbia, Canada
\\$^1${firstname.lastname}@inria.fr, $^2$nmisyats@ece.ubc.ca
\country{}
}
%

\renewcommand{\shortauthors}{M. Traiola et al.}

\begin{abstract}
Error-tolerant applications, such as multimedia processing, machine learning, signal processing, and scientific computing can still produce satisfactory results under approximate computation. Approximate computing (AxC) is a well-established paradigm that improves efficiency by judiciously reducing computation quality, and has been widely applied to arithmetic circuits by modifying their logic behavior. However, depending on the approximation, arithmetic properties, such as commutativity, may no longer hold, causing error accumulation and making the output depend on the operand order. In this work, we show that controlling operand order in non-commutative approximate circuits can significantly reduce computational errors.
We propose SWAPPER, a lightweight approach that drastically reduces approximation error by dynamically reordering input operands using only a single bit for the decision. To identify the most effective bit for swapping, we propose a framework applicable at different granularities, achieving large error reductions and significant accuracy gains. We evaluate SWAPPER at both component and application levels: at the component level, it reduces the mean absolute error by up to 50\%, and at the application level it achieves up to 90\% error reduction for AxBench suite and 66.7\% top-1 improvement on a Vision Transformer.
\vspace{-5pt}
\end{abstract}

\keywords{Approximate Computing, Approximate Circuits, Non-commutative, Online Error Reduction, Dynamic Swap
\vspace{-5pt}}

\maketitle

\thispagestyle{firstpage}
\section{Introduction}
Error-tolerant applications, including multimedia processing, machine learning, signal processing, and scientific computing, can deliver acceptable results even with inexact computations. Approximate computing (AxC) exploits this tolerance by trading exactness for reductions in area, energy, and delay through design simplifications at various levels~\cite{mittal-axc,henkel2022}. For instance, replacing 32-bit floating-point with low-precision ($16$-, $12$-, or $8$-bit) integer formats reduces memory footprint and enables simpler, lower-power arithmetic units. Logic behavior can also be simplified, e.g., by removing gates, yielding Approximate Integrated Circuits (AxICs)~\cite{Jie13,Jie14,Jie15,Jie16,Jie18,JieSurvey20}.

However, this circuit simplification often results in the loss of mathematical properties, such as commutativity, compared to the exact operator~\cite{Lombardi-commutative-nanoarch21,lombardiCommutative24}.
Typical AxIC approaches, such as approximate logic synthesis~\cite{ALSReda-18,ALS-Chang20, ALS-Chang22,DSEJorge22,ALS-Chang23,DSEMario23} and evolutionary design~\cite{Vasicek15,evoapprox,Vasicek17,Vasicek18,Vasicek20}, lead to both commutative and non-commutative designs.
Recent work has extended evolutionary design with constraints that enforce commutativity in AxIC multipliers and adders~\cite{Vasicek-commutativeDesign24}, showing that commutative AxICs can use a similar number of gates as non-commutative ones with comparable Worst-Case absolute arithmetic Error (WCE).
To improve the accuracy reduction due to AxC, existing approaches  typically apply error-correction techniques~\cite{correctionICCD13,Jie14,correctionISCAS16,correctionGLSVLSIS16,correctionISQED21,correctionESL22}.
AxICs with double-sided error distributions are used to mitigate error accumulation in complex computations~\cite{Jie-doubleside-22,Jie-doubleside-24}.
Nonetheless, non-commutative AxICs have properties that offer complementary opportunities for accuracy improvement.
It has been shown that non-commutative AxIC adders exhibit different error profiles depending on operand order~\cite{lombardiCommutative24}. However, these properties have not been explored yet for improving the accuracy reduction.

Building on this insight, we propose SWAPPER, an approach that dynamically determines the order of operands during execution to efficiently and proactively reduce the approximation error of non-commutative AxICs.
Predicting whether swapping the operands will be beneficial for the computation is not straightforward.
To address this problem, we propose:
\begin{enumerate*}[label=(\roman*)]
\item a lightweight mechanism that uses only one bit to decide on operand swapping, and
\item an exploration framework to guide the decision-making process for the operand order.
\end{enumerate*}
The framework can be applied at different granularities: i) component level, taking into account only the specific AxICs characteristics, and ii) application level, enhancing the exploration with the application characteristics. 
Extensive experimental results evaluate our methodology using the EvoApproxLib multipliers~\cite{evoapprox}, the AxBench CPU benchmark suite~\cite{Yaz16}, and a Vision Transformer (ViT)~\cite{ViT} as case studies. Results at both component and application levels show substantial error reductions: up to 50\% in Mean Absolute Error at component level, up to 90\% for AxBench suite at the application level, and 66.7\% accuracy improvement for ViT.

\section{Proposed methodology}
\label{sec:proposed}
The proposed methodology dynamically decides the operand swapping for non-commutative AxICs. To remain lightweight, it uses only a single bit for this decision. Before execution, a tuning phase applies an exploration framework to identify the most effective bit to be used for the run-time decision.

\noindent{\bf Tuning phase:}
An exhaustive exploration of all operand bits is performed, i.e., the error (based on a selected metric) is calculated when each bit $i \in \{0, ..., M\}$ of both A and B is selected as the bit to be used for the run-time decision, where $M$ is the size of the input operands. %
The result is the combination -- the operand ($A$ or $B$), the bit position ($\{0, ..., M\}$), and the bit value ($0$ or $1$) -- that gives the minimum error.
As the error depends on the expected distribution of the operator inputs, SWAPPER provides two main exploration granularity levels for the operand values during the tuning phase:
\begin{itemize}[leftmargin=*]
    \item \emph{Component level:} all possible value combinations that can be represented with the bit-width ($M$ bits) are exhaustively used for operands $A$ and $B$, because no additional information about the application and inputs is available at this level. The tuning needs to be done once, as all possible operand values are considered. 
     \item \emph{Application level:} only a subset of values is used for operands $A$ and $B$,  given by the application execution using representative inputs. The tuning is performed for each application. 
\end{itemize}
Note that, by limiting the decision to a single bit, there is no need to explore combinations of multiple bits, significantly reducing the exploration complexity. Furthermore, the same bit is used to make swapping decisions on all multiplications within an application.

The framework currently supports common error metrics at component level, such as  \textit{Mean Absolute Error} (MAE), \textit{Worst Case Error} (WCE), \textit{Average Relative Error} (ARE), \textit{Mean Squared Error} (MSE), and \textit{Error Probability} (EP)~\cite{metrics,evoapprox}:
\vspace{-\baselineskip}

 {\small
\begin{alignat*}{2}
\text{MAE} &= \frac{1}{N}\sum_{i=1}^{N}|O_i^{\text{approx}}-O_i^{\text{precise}}|
&\qquad&
\text{WCE} = \max_{i\in\textit{Inputs}}|O_i^{\text{approx}}-O_i^{\text{precise}}|\\
\text{ARE} &= \frac{1}{N}\sum_{i=1}^{N}
\frac{|O_i^{\text{approx}}-O_i^{\text{precise}}|}{|O_i^{\text{precise}}|}
&&
\text{MSE} = \frac{1}{N}\sum_{i=1}^{N}|O_i^{\text{approx}}-O_i^{\text{precise}}|^2
\end{alignat*}
\begin{equation*}
\text{EP} = \frac{1}{N} \sum_{i=1}^{N} \left\{
\begin{array}{ll}
1, & \text{if } O_{i}^{\text{approx}} \neq O_{i}^{\text{precise}} \\
0, & \text{otherwise}
\end{array}
\right.
\end{equation*}
}

\noindent where $N$ is the number of inputs, $O_{i}^{\text{approx}}$ is the approximate output, $O_{i}^{\text{precise}}$ the precise output, and $\left\vert O_{i}^{\text{approx}}- O_{i}^{\text{precise}}\right\vert$ the absolute error. The framework can be easily extended with other metrics. Application-level error metrics are determined based on the specific application requirements, as illustrated in the experimental section.

\noindent{\bf Execution:} The result of the tuning phase is used by the run-time SWAPPER mechanism to determine the operand order during execution. SWAPPER can be implemented in software or hardware, depending on the platform and design choices; Experimental section describes and evaluates one hardware implementation.

\noindent{\textit{\textbf{Illustrative example:}}}\label{sec:illustrativeExample}
We illustrate how SWAPPER exploits the properties of non-commutative AxICs with an example. For clarity, we demonstrate this phenomenon using two AxICs that exhibit large errors across the input range. Consider two 8-bit multipliers from EvoApproxLib~\cite{evoapprox}, with $A$ and $B$ as primary inputs, and $x$ and $y$ as generic input values:
\begin{enumerate*}[label=(\roman*)]
\item \texttt{mul8u\_17MJ} satisfies the commutative property (i.e., $x\cdot y=y\cdot x$),  and 
\item \texttt{mul8u\_17MN} does not satisfy the commutative property (i.e., $x\cdot y\neq x\cdot y$).
\end{enumerate*}
Swapping the operands $x$ and $y$ for \texttt{mul8u\_17MN} leads to a different error $E$ (i.e., $E_{x\cdot y}\neq E_{y\cdot x}$).

\begin{figure}[tp]
   \centering
\includegraphics[width=\columnwidth]{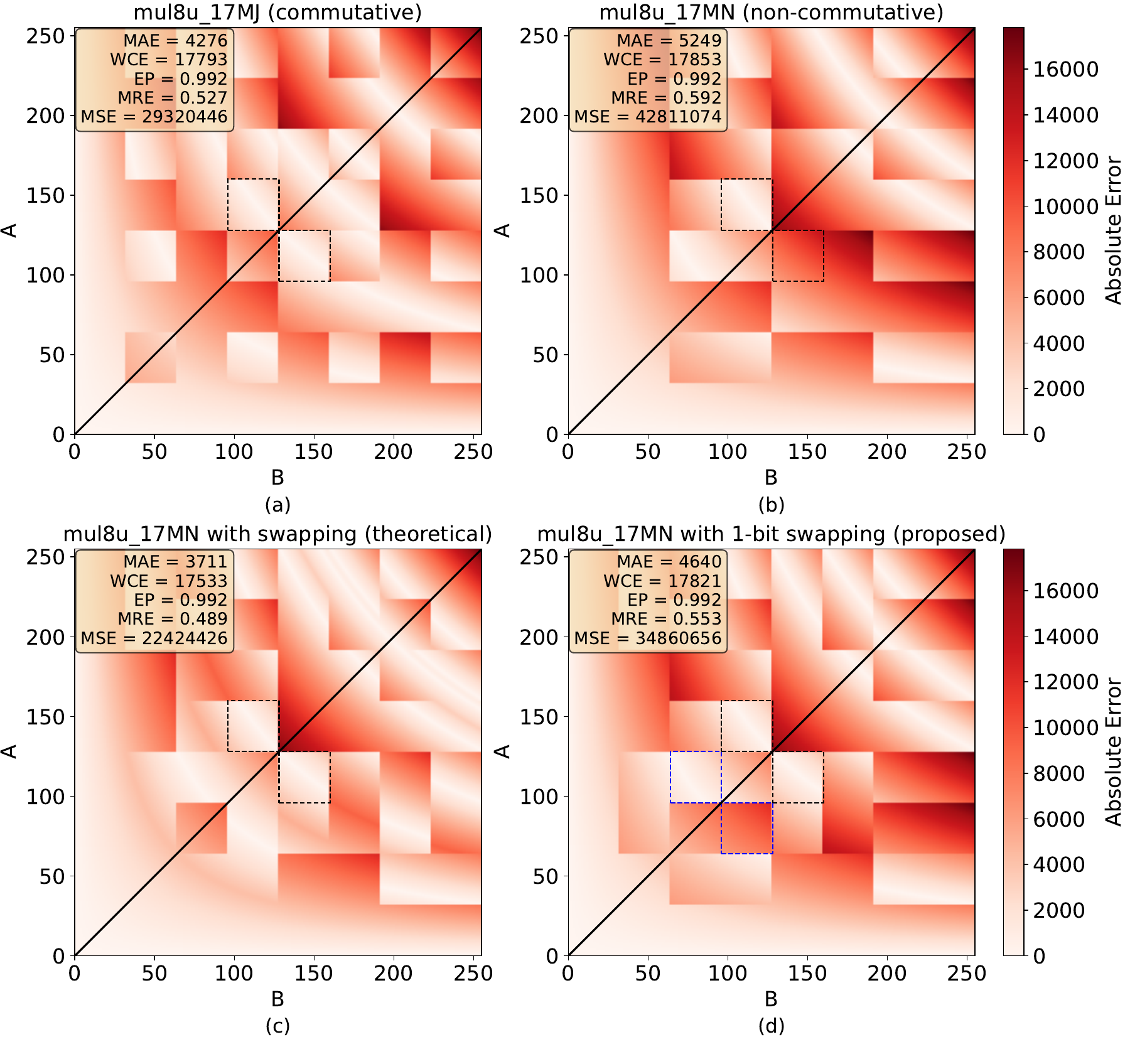}
\vspace{-25pt}
    \caption{Error profile of two $8$-bit unsigned multipliers (EvoapproxLib~\cite{evoapprox}): (a) \texttt{mul8u\_17MJ} (commutative, $A\cdot B=B \cdot A$) and \texttt{mul8u\_17MN} (non-commutative, $A\cdot B\neq B\cdot A$) (b) without swap and (d) with SWAPPER. (c) is the theoretical scenario.}
\label{fig:motivating-example}
\vspace{-8pt}
\end{figure}

Figure~\ref{fig:motivating-example} shows the error profile -- visualized as a heat map where each pixel is the absolute error -- and the standard component-level error metrics for (a) the commutative \texttt{mul8u\_17MJ} and (b) the non-commutative \texttt{mul8u\_17MN}.
Figure~\ref{fig:motivating-example}(a) is symmetric with respect to the diagonal $A=B$ (black line), as is always the case for commutative AxICs, where swapping $A=x$, $B=y$ (with $x \neq y$) does not change the output error.
In contrast, Figure~\ref{fig:motivating-example}(b) shows that this symmetry does not hold for non-commutative AxICs. For instance, the color patterns within the dashed black squares in Figure~\ref{fig:motivating-example}(b) differ substantially, whereas in Figure~\ref{fig:motivating-example}(a) they are symmetric due to commutativity.
The error profile of the non-commutative multiplier \texttt{mul8u\_17MN} changes substantially when the operands are swapped. Theoretically, an oracle that always selects the operand order yielding the smallest possible error reaches the profile of Figure~\ref{fig:motivating-example}(c), improving MAE by 29.3\%, WCE by 1.8\%, ARE by 17.5\%, and MSE by 47.6\%. However, implementing such an oracle is hardly feasible, as it requires to use different decision bits each time depending on the input values.
Figure~\ref{fig:motivating-example}(d) shows the error profile of \texttt{mul8u\_17MN} with SWAPPER applied at the component level. Swapping operands using only a single bit (only when the 6$^\text{th}$ bit of operand B is 0) improves MAE by 11.6\%, WCE by 0.2\%, ARE by 6.6\%, and MSE by 18.6\%. Overall, SWAPPER improves the symmetry in the error profile (black dashed squares), but not in all cases (blue dashed squares), as it uses only a single decision bit, significantly reducing the exploration and run-time decision complexity. 

\begin{table*}[tp]
\setlength{\tabcolsep}{2pt}
\centering
\caption{Original MAE, MAE reduction obtained with SWAPPER, and theoretical maximum MAE reduction for non-commutative $8$-bit and $16$-bit unsigned (\texttt{8u\_}, \texttt{16u\_}) and signed (\texttt{8s\_}, \texttt{16s\_}) multipliers from EvoApproxLib~\cite{evoapprox}.}
\vspace{-10pt}
\label{tab:mul}
\resizebox{\textwidth}{!}{%
\begin{tabular}{l*{14}{c}}
\multicolumn{9}{l}{%
    \multirow{2}{*}{\barmulgraphic}%
} &
\multicolumn{3}{||r}{\textbf{Commutative Ax multiplier}} &
\textbf{\texttt{16s\_HF7}} &
\textbf{\texttt{16s\_HG8}} &
\textbf{\texttt{16s\_HFB}} \\

&&&&&&&&&
\multicolumn{3}{||r}{Original} &
13653.00 &
38343.00 &
87012.00 \\
\hline\hline
\textbf{Ax multiplier} & 
\textbf{\texttt{8u\_12YX}} & \textbf{\texttt{8u\_27Y}} & \textbf{\texttt{8u\_4X5}} & \textbf{\texttt{8u\_2V0}} & \textbf{\texttt{8u\_ZDF}} & \textbf{\texttt{8u\_DG8}} & \textbf{\texttt{8u\_C67}} & \textbf{\texttt{8u\_R92}} & \textbf{\texttt{8u\_12KA}} & \textbf{\texttt{8u\_GTR}} & \textbf{\texttt{8u\_874}} & \textbf{\texttt{8u\_0AB}} & \textbf{\texttt{8u\_NLX}} & \textbf{\texttt{8u\_XFM}} \\ 
Original  & 0.50 & 0.88 & 0.91 & 1.00 & 4.40 & 5.18 & 8.66 & 11.37 & 12.00 & 29.43 & 37.04 & 37.07 & 42.29 & 43.81 \\
SWAPPER  & \cellcolor{mycolorhigh!25!mycolorlow} 25.00\% & \cellcolor{mycolorhigh!14!mycolorlow} 14.29\% & \cellcolor{mycolorhigh!20!mycolorlow} 20.69\% & \cellcolor{mycolorhigh!20!mycolorlow} 20.31\% & \cellcolor{mycolorhigh!11!mycolorlow} 11.00\% & \cellcolor{mycolorhigh!14!mycolorlow} 14.03\% & \cellcolor{mycolorhigh!14!mycolorlow} 14.30\% & \cellcolor{mycolorhigh!7!mycolorlow} 7.03\% & \cellcolor{mycolorhigh!50!mycolorlow} 50.00\% & \cellcolor{mycolorhigh!12!mycolorlow} 12.27\% & \cellcolor{mycolorhigh!39!mycolorlow} 39.81\% & \cellcolor{mycolorhigh!1!mycolorlow} 1.75\% & \cellcolor{mycolorhigh!3!mycolorlow} 3.84\% & \cellcolor{mycolorhigh!8!mycolorlow} 8.57\% \\
Theoretical  & \cellcolor{gray!20} 37.50\% & \cellcolor{gray!20} 28.57\% & \cellcolor{gray!20} 27.59\% & \cellcolor{gray!20} 55.86\% & \cellcolor{gray!20} 28.22\% & \cellcolor{gray!20} 38.91\% & \cellcolor{gray!20} 47.03\% & \cellcolor{gray!20} 38.91\% & \cellcolor{gray!20} 92.71\% & \cellcolor{gray!20} 43.30\% & \cellcolor{gray!20} 88.10\% & \cellcolor{gray!20} 30.56\% & \cellcolor{gray!20} 24.66\% & \cellcolor{gray!20} 26.94\% \\
\hline\hline
\textbf{Ax multiplier} & 
\textbf{\texttt{8u\_8U3}} & \textbf{\texttt{8u\_197B}} & \textbf{\texttt{8u\_19XF}} & \textbf{\texttt{8u\_L93}} & \textbf{\texttt{8u\_BG1}} & \textbf{\texttt{8u\_2NDH}} & \textbf{\texttt{8u\_17C8}} & \textbf{\texttt{8u\_1DMU}} & \textbf{\texttt{8u\_17R6}} & \textbf{\texttt{8u\_4TF}} & \textbf{\texttt{8u\_18UH}} & \textbf{\texttt{8u\_GJM}} & \textbf{\texttt{8u\_T83}} & \textbf{\texttt{8u\_1A0M}} \\ 
Original  & 87.25 & 119.02 & 148.15 & 157.01 & 284.31 & 285.47 & 370.05 & 426.28 & 441.61 & 731.44 & 773.11 & 1011.25 & 1409.42 & 1888.59 \\
SWAPPER  & \cellcolor{mycolorhigh!33!mycolorlow} 33.13\% & \cellcolor{mycolorhigh!4!mycolorlow} 4.89\% & \cellcolor{mycolorhigh!5!mycolorlow} 5.25\% & \cellcolor{mycolorhigh!5!mycolorlow} 5.29\% & \cellcolor{mycolorhigh!37!mycolorlow} 37.66\% & \cellcolor{mycolorhigh!8!mycolorlow} 8.83\% & \cellcolor{mycolorhigh!2!mycolorlow} 2.37\% & \cellcolor{mycolorhigh!24!mycolorlow} 24.74\% & \cellcolor{mycolorhigh!1!mycolorlow} 1.28\% & \cellcolor{mycolorhigh!43!mycolorlow} 43.17\% & \cellcolor{mycolorhigh!5!mycolorlow} 5.13\% & \cellcolor{mycolorhigh!29!mycolorlow} 29.66\% & \cellcolor{mycolorhigh!2!mycolorlow} 2.25\% & \cellcolor{mycolorhigh!2!mycolorlow} 2.92\% \\
Theoretical  & \cellcolor{gray!20} 83.62\% & \cellcolor{gray!20} 26.95\% & \cellcolor{gray!20} 39.74\% & \cellcolor{gray!20} 36.18\% & \cellcolor{gray!20} 73.85\% & \cellcolor{gray!20} 34.16\% & \cellcolor{gray!20} 17.16\% & \cellcolor{gray!20} 60.73\% & \cellcolor{gray!20} 32.30\% & \cellcolor{gray!20} 81.68\% & \cellcolor{gray!20} 17.10\% & \cellcolor{gray!20} 56.30\% & \cellcolor{gray!20} 14.74\% & \cellcolor{gray!20} 23.28\% \\
\hline\hline
\textbf{Ax multiplier} & 
\textbf{\texttt{8u\_Z9D}} & \textbf{\texttt{8u\_R36}} & \textbf{\texttt{8u\_17MN}} & \textbf{\texttt{8u\_1SX}} & \textbf{\texttt{8s\_1KVM}} & \textbf{\texttt{8s\_1KVP}} & \textbf{\texttt{8s\_1KVQ}} & \textbf{\texttt{8s\_1KX5}} & \textbf{\texttt{8s\_1KXF}} & \textbf{\texttt{8s\_1L12}} & \textbf{\texttt{12u\_22U}} & \textbf{\texttt{12u\_10C}} & \textbf{\texttt{12u\_2CC}} & \textbf{\texttt{12u\_2CD}} \\ 
Original  & 3169.34 & 3828.22 & 5249.49 & 6362.00 & 32.00 & 33.28 & 36.54 & 101.28 & 224.00 & 2016.00 & 0.88 & 970.57 & 1023.75 & 1024.25 \\
SWAPPER  & \cellcolor{mycolorhigh!0!mycolorlow} 0.07\% & \cellcolor{mycolorhigh!6!mycolorlow} 6.21\% & \cellcolor{mycolorhigh!11!mycolorlow} 11.61\% & \cellcolor{mycolorhigh!2!mycolorlow} 2.69\% & \cellcolor{mycolorhigh!50!mycolorlow} 50.00\% & \cellcolor{mycolorhigh!45!mycolorlow} 45.08\% & \cellcolor{mycolorhigh!35!mycolorlow} 35.75\% & \cellcolor{mycolorhigh!28!mycolorlow} 28.43\% & \cellcolor{mycolorhigh!28!mycolorlow} 28.57\% & \cellcolor{mycolorhigh!25!mycolorlow} 25.40\% & \cellcolor{mycolorhigh!7!mycolorlow} 7.14\% & \cellcolor{mycolorhigh!5!mycolorlow} 5.23\% & \cellcolor{mycolorhigh!49!mycolorlow} 49.99\% & \cellcolor{mycolorhigh!49!mycolorlow} 49.94\% \\
Theoretical  & \cellcolor{gray!20} 0.13\% & \cellcolor{gray!20} 9.64\% & \cellcolor{gray!20} 29.30\% & \cellcolor{gray!20} 9.59\% & \cellcolor{gray!20} 66.66\% & \cellcolor{gray!20} 61.29\% & \cellcolor{gray!20} 51.85\% & \cellcolor{gray!20} 49.81\% & \cellcolor{gray!20} 51.72\% & \cellcolor{gray!20} 47.51\% & \cellcolor{gray!20} 7.14\% & \cellcolor{gray!20} 39.33\% & \cellcolor{gray!20} 66.66\% & \cellcolor{gray!20} 66.60\% \\
\hline\hline
\textbf{Ax multiplier} & 
\textbf{\texttt{12u\_2CJ}} & \textbf{\texttt{12u\_2D0}} & \textbf{\texttt{12u\_2DL}} & \textbf{\texttt{12u\_2DA}} & \textbf{\texttt{12u\_2DV}} & \textbf{\texttt{12u\_2EF}} & \textbf{\texttt{12u\_2F1}} & \textbf{\texttt{12s\_2QD}} & \textbf{\texttt{12s\_2QE}} & \textbf{\texttt{12s\_2QH}} & \textbf{\texttt{12s\_2R5}} & \textbf{\texttt{12s\_2RP}} & \textbf{\texttt{12s\_2TE}} & \textbf{\texttt{16u\_3CF}} \\ 
Original  & 1184.25 & 3071.25 & 7166.25 & 12288.25 & 15360.25 & 30720.25 & 88064.25 & 512.00 & 512.25 & 524.38 & 1540.31 & 3584.00 & 32256.00 & 1.34 \\
SWAPPER  & \cellcolor{mycolorhigh!41!mycolorlow} 41.88\% & \cellcolor{mycolorhigh!33!mycolorlow} 33.31\% & \cellcolor{mycolorhigh!28!mycolorlow} 28.52\% & \cellcolor{mycolorhigh!4!mycolorlow} 4.17\% & \cellcolor{mycolorhigh!9!mycolorlow} 10.00\% & \cellcolor{mycolorhigh!9!mycolorlow} 10.00\% & \cellcolor{mycolorhigh!4!mycolorlow} 4.65\% & \cellcolor{mycolorhigh!50!mycolorlow} 50.00\% & \cellcolor{mycolorhigh!49!mycolorlow} 49.93\% & \cellcolor{mycolorhigh!47!mycolorlow} 47.30\% & \cellcolor{mycolorhigh!33!mycolorlow} 33.05\% & \cellcolor{mycolorhigh!28!mycolorlow} 28.57\% & \cellcolor{mycolorhigh!25!mycolorlow} 25.40\% & \cellcolor{mycolorhigh!16!mycolorlow} 16.28\% \\
Theoretical  & \cellcolor{gray!20} 56.28\% & \cellcolor{gray!20} 56.00\% & \cellcolor{gray!20} 51.69\% & \cellcolor{gray!20} 6.25\% & \cellcolor{gray!20} 17.29\% & \cellcolor{gray!20} 17.29\% & \cellcolor{gray!20} 6.98\% & \cellcolor{gray!20} 66.67\% & \cellcolor{gray!20} 66.59\% & \cellcolor{gray!20} 63.62\% & \cellcolor{gray!20} 55.60\% & \cellcolor{gray!20} 51.74\% & \cellcolor{gray!20} 48.56\% & \cellcolor{gray!20} 22.09\% \\
\hline\hline
\textbf{Ax multiplier} & 
\textbf{\texttt{16u\_7EL}} & \textbf{\texttt{16u\_9GU}} & \textbf{\texttt{16u\_3H1}} & \textbf{\texttt{16u\_7G2}} & \textbf{\texttt{16u\_9M4}} & \textbf{\texttt{16u\_9MP}} & \textbf{\texttt{16u\_1K3}} & \textbf{\texttt{16u\_3PM}} & \textbf{\texttt{16u\_7L2}} & \textbf{\texttt{16u\_9YG}} & \textbf{\texttt{16u\_A1M}} & \textbf{\texttt{16u\_7QP}} & \textbf{\texttt{16u\_2KD}} & \textbf{\texttt{16u\_21M}} \\ 
Original  & 2.93 & 4.88 & 12.41 & 24.53 & 37.07 & 41.41 & 64.19 & 129.17 & 194.71 & 370.05 & 458.82 & 1136.79 & 3230.19 & 12373.78 \\
SWAPPER  & \cellcolor{mycolorhigh!8!mycolorlow} 8.53\% & \cellcolor{mycolorhigh!5!mycolorlow} 5.77\% & \cellcolor{mycolorhigh!4!mycolorlow} 4.75\% & \cellcolor{mycolorhigh!6!mycolorlow} 6.16\% & \cellcolor{mycolorhigh!1!mycolorlow} 1.75\% & \cellcolor{mycolorhigh!6!mycolorlow} 7.00\% & \cellcolor{mycolorhigh!49!mycolorlow} 49.85\% & \cellcolor{mycolorhigh!4!mycolorlow} 4.19\% & \cellcolor{mycolorhigh!4!mycolorlow} 4.33\% & \cellcolor{mycolorhigh!2!mycolorlow} 2.37\% & \cellcolor{mycolorhigh!1!mycolorlow} 1.88\% & \cellcolor{mycolorhigh!3!mycolorlow} 3.13\% & \cellcolor{mycolorhigh!1!mycolorlow} 1.68\% & \cellcolor{mycolorhigh!16!mycolorlow} 16.57\% \\
Theoretical  & \cellcolor{gray!20} 37.13\% & \cellcolor{gray!20} 24.68\% & \cellcolor{gray!20} 40.15\% & \cellcolor{gray!20} 27.52\% & \cellcolor{gray!20} 30.56\% & \cellcolor{gray!20} 39.50\% & \cellcolor{gray!20} 74.77\% & \cellcolor{gray!20} 39.35\% & \cellcolor{gray!20} 26.76\% & \cellcolor{gray!20} 17.16\% & \cellcolor{gray!20} 35.37\% & \cellcolor{gray!20} 34.08\% & \cellcolor{gray!20} 30.88\% & \cellcolor{gray!20} 24.96\% \\
\hline\hline
\textbf{Ax multiplier} & 
\textbf{\texttt{16u\_4UG}} & \textbf{\texttt{16u\_4XB}} & \textbf{\texttt{16u\_G4R}} & \textbf{\texttt{16u\_5A2}} & \textbf{\texttt{16u\_60G}} & \textbf{\texttt{16u\_G87}} & \textbf{\texttt{16u\_G88}} & \textbf{\texttt{16u\_GAU}} & \textbf{\texttt{16u\_GPV}} & \textbf{\texttt{16s\_GQU}} & \textbf{\texttt{16s\_GQV}} & \textbf{\texttt{16s\_GRU}} & \textbf{\texttt{16s\_GSM}} & \textbf{\texttt{16s\_GV3}} \\ 
Original  & 31363.63 & 54059.25 & 114688.25 & 132023.36 & 463045.66 & 4423680.00 & 6782976.00 & 75464704.00 & 668168640.00 & 8192.00 & 8192.25 & 24580.25 & 57344.00 & 516096.00 \\
SWAPPER  & \cellcolor{mycolorhigh!1!mycolorlow} 1.26\% & \cellcolor{mycolorhigh!3!mycolorlow} 3.15\% & \cellcolor{mycolorhigh!28!mycolorlow} 28.57\% & \cellcolor{mycolorhigh!3!mycolorlow} 3.79\% & \cellcolor{mycolorhigh!2!mycolorlow} 2.65\% & \cellcolor{mycolorhigh!8!mycolorlow} 8.89\% & \cellcolor{mycolorhigh!3!mycolorlow} 3.86\% & \cellcolor{mycolorhigh!5!mycolorlow} 5.56\% & \cellcolor{mycolorhigh!10!mycolorlow} 10.04\% & \cellcolor{mycolorhigh!49!mycolorlow} 50.00\% & \cellcolor{mycolorhigh!49!mycolorlow} 50.00\% & \cellcolor{mycolorhigh!33!mycolorlow} 33.32\% & \cellcolor{mycolorhigh!28!mycolorlow} 28.57\% & \cellcolor{mycolorhigh!25!mycolorlow} 25.40\% \\
Theoretical  & \cellcolor{gray!20} 29.29\% & \cellcolor{gray!20} 38.97\% & \cellcolor{gray!20} 51.73\% & \cellcolor{gray!20} 33.53\% & \cellcolor{gray!20} 33.32\% & \cellcolor{gray!20} 15.37\% & \cellcolor{gray!20} 5.80\% & \cellcolor{gray!20} 8.34\% & \cellcolor{gray!20} 15.07\% & \cellcolor{gray!20} 66.67\% & \cellcolor{gray!20} 66.66\% & \cellcolor{gray!20} 55.99\% & \cellcolor{gray!20} 51.74\% & \cellcolor{gray!20} 48.56\% \\
\hline\hline
\end{tabular}
}
\end{table*}
\section{Experimental evaluation}

We experimentally evaluate SWAPPER at both component and application level. We consider as a case study the complete set (C++ models) of $16$-bit, $12$-bit, and $8$-bit multipliers from EvoApproxLib~\cite{evoapprox}, unsigned and signed, the CPU benchmarks from AxBench benchmark suite~\cite{Yaz16}, and a Vision Transformer~\cite{ViT}, to demonstrate SWAPPER's applicability and results with a larger application.
We focus on multiplication as it leads to more significant errors when approximated. However, the methodology is applicable to other arithmetic operators with the same properties, e.g., addition.
We evaluate the proposed approach by comparing SWAPPER with the scenario where no dynamic swapping is applied (NoSwap). For every case, we detail the error metrics used for the comparison.
We initially present the evaluation results considering the error at the component level (output of each multiplier of EvoApproxLib), and then at the application level (output of each benchmark).

\subsection{Component level}
To perform the tuning phase at the component level, an M-bit input circuit, with $A$ and $B$ as primary inputs, is exhaustively stimulated (i.e., $2^{2M}$ input combinations) while dynamically deciding whether to swap or not, depending on one bit of $A$ or $B$ operand. This exploration is performed for all bits of $A$ and $B$ (i.e., $2M$) and for both reference values (i.e., the decision bit equal to $0$ or $1$). This leads to a total of $4M \cdot 2^{2M}$ circuit stimulations. 
 The experiments were performed on  AMD EPYC 7443 processor running at 2.85GHz. The maximum measured time for the tuning phase was around 3 hours for the largest circuits, i.e., 16-bit multipliers, for which $64\cdot 2^{32}$ stimulations were performed without thread parallelization.

Table~\ref{tab:mul} presents the MAE reduction for the complete set of multipliers from EvoApproxLib. The ``original'' rows report the MAE value of the original multipliers, as presented in the library. The ``SWAPPER'' rows report the best relative MAE reduction ($\%$) obtained when applying our approach.
The ``Theoretical'' rows report the maximum possible relative MAE reduction obtained by choosing, for each multiplication, the order ($A \cdot B$ or $B \cdot A$) that yields the lower absolute error. This corresponds to an oracle setting, as in the example in Section~\ref{sec:illustrativeExample} and Figure~\ref{fig:motivating-example}(c). We report these values to show how closely SWAPPER can approach this oracle using only one bit for dynamic operand swapping.
Overall, using a single control bit to select the operand order can reduce the output error of each multiplier from 1.26\% to 50\%, depending on the multiplier design. For comparison, we also report results for 16-bit signed commutative multipliers, which have an MAE similar to the non-commutative ones, but lack asymmetric properties that could be exploited for additional error reduction.

\begin{table*}[htp!]%
\captionsetup[subfigure]{skip=0.5pt}
  \centering
 \caption{AxBench error comparison: Original, FxP, cummutative and non-commutative \texttt{16s\_} multipliers from EvoApproxLib~\cite{evoapprox}.}
  \label{tbl:all-approx-table}%
\vspace{-10pt}
\begin{minipage}[t]{0.49\textwidth}
\centering
\setlength{\tabcolsep}{1.5pt}
\centering
\resizebox{1.05\columnwidth}{!}{%
\begin{tabular}{|c|c|c|c|c|c|c|c|}
\multicolumn{8}{c}{\barexpsym%
}\\
\hline
  & \texttt{Jmeint} & \texttt{Blackscholes} & \texttt{FFT} & \texttt{Inversek2j} & \texttt{Kmeans} & \texttt{Sobel} & \texttt{Jpeg} \\
\cline{2-8}
& {\bf Miss Rate} & \multicolumn{3}{c|}{{\bf ARE}} & \multicolumn{3}{c|}{{\bf Avg. SSIM}} \\
\hline
\textbf{Original} & 
\cellcolor{green!100.0000!orange!40}0\% & 
\cellcolor{green!100.0000!orange!40}2.4\% & 
\cellcolor{green!100.0000!orange!40}0\% & 
\cellcolor{green!100.0000!orange!40}0.0046\% & 
\cellcolor{green!100.0000!orange!40}1 & 
\cellcolor{green!100.0000!orange!40}1 & 
\cellcolor{green!100.0000!orange!40}1\\
\hline
\textbf{FxP} & 
\cellcolor{green!83.4000!orange!40}8.3\% & 
\cellcolor{green!93.6475!orange!40}5.5\% & 
\cellcolor{green!87.5000!orange!40}6.25\% & 
\cellcolor{green!99.0892!orange!40}0.46\% & 
\cellcolor{green!99.6000!orange!40}0.998 & 
\cellcolor{green!99.6000!orange!40}0.998 & \\\hline\hline
\multicolumn{8}{|c|}{\textbf{Commutative AxIC multipliers}} \\
\hline
HF7 & 
\cellcolor{green!30.5600!orange!40}34.72\% & 
\cellcolor{orange!4.6200!red!40}97.69\% & 
\cellcolor{orange!17.9800!red!40}91.01\% & 
\cellcolor{orange!40.8000!red!40}79.60\% & 
\cellcolor{orange!90.1750!red!40}0.4509 & 
\cellcolor{orange!6.3449!red!40}0.317 & 
\cellcolor{green!98.6542!orange!40}0.9933 \\
\cline{1-7}
HG8 & 
\cellcolor{green!35.0400!orange!40}32.48\% & 
\cellcolor{orange!4.1000!red!40}97.95\% & 
\cellcolor{orange!3.0400!red!40}98.48\% & 
\cellcolor{orange!45.7400!red!40}77.13\% & 
\cellcolor{orange!90.1750!red!40}0.4509 & 
\cellcolor{orange!6.2475!red!40}0.312 & 
\cellcolor{green!96.3267!orange!40}0.9816 \\
\cline{1-7}
HFB & 
\cellcolor{green!37.5000!orange!40}31.25\% & 
\cellcolor{orange!6.4800!red!40}96.76\% & 
\cellcolor{orange!7.4800!red!40}96.26\% & 
\cellcolor{orange!46.8600!red!40}76.57\% & 
\cellcolor{orange!90.1750!red!40}0.4509 & 
\cellcolor{orange!6.2346!red!40}0.312 & 
\cellcolor{green!91.1820!orange!40}0.9559 \\
\cline{1-8}
\end{tabular}
}
\end{minipage}
\begin{minipage}[t]{0.49\textwidth}
\centering
\setlength{\tabcolsep}{2pt}
\centering
\small

\resizebox{0.88\columnwidth}{!}{%
\begin{tabular}{|c|c|c|c|c|}
\multicolumn{1}{l}{\texttt{Jmeint}} & \multicolumn{2}{c}{ \textbf{Miss rate} (lower is better) } &\multicolumn{2}{r}{\barexp%
}\\
\hline
\textbf{Non-cummutative} & \multicolumn{1}{c|}{\multirow{2}{*}{\textbf{NoSwap}}} & \multicolumn{2}{c|}{\textbf{SWAPPER}} & \multirow{2}{*}{\textbf{Theor.}} \\ \cline{3-4}
\textbf{AxIC multipliers} & \multicolumn{1}{c|}{} & \multicolumn{1}{c|}{\textbf{Comp.}} & \multicolumn{1}{c|}{\textbf{App.}} &  \\ \hline
GQU & \multicolumn{1}{c|}{\cellcolor{green!33.5685!orange!40}33.22\%} & \multicolumn{1}{c|}{\cellcolor{green!41.8984!orange!40}29.05\%} & \multicolumn{1}{c|}{\cellcolor{green!98.7038!orange!40}0.65\%} & \cellcolor{green!98.5200!orange!40}0.74\% \\ \hline
GQV & \multicolumn{1}{c|}{\cellcolor{green!33.5705!orange!40}33.21\%} & \multicolumn{1}{c|}{\cellcolor{green!41.5560!orange!40}29.22\%} & \multicolumn{1}{c|}{\cellcolor{green!97.2024!orange!40}1.40\%} & \cellcolor{green!97.1552!orange!40}1.42\% \\ \hline
GRU & \multicolumn{1}{c|}{\cellcolor{green!36.7752!orange!40}31.61\%} & \multicolumn{1}{c|}{\cellcolor{green!38.9305!orange!40}30.53\%} & \multicolumn{1}{c|}{\cellcolor{green!79.0426!orange!40}10.48\%} & \cellcolor{green!79.0935!orange!40}10.45\% \\ \hline
GSM & \multicolumn{1}{c|}{\cellcolor{green!38.6271!orange!40}30.69\%} & \multicolumn{1}{c|}{\cellcolor{green!41.6937!orange!40}29.15\%} & \multicolumn{1}{c|}{\cellcolor{green!95.8958!orange!40}2.05\%} & \cellcolor{green!94.9119!orange!40}2.54\% \\ \hline
GV3 & \multicolumn{1}{c|}{\cellcolor{green!41.3307!orange!40}29.33\%} & \multicolumn{1}{c|}{\cellcolor{green!42.4485!orange!40}28.78\%} & \multicolumn{1}{c|}{\cellcolor{green!85.7251!orange!40}7.14\%} & \cellcolor{green!80.4467!orange!40}9.78\% \\ \hline
\multicolumn{1}{|l|}{\textbf{\scriptsize Gain w.r.t. NoSwap}} & \multicolumn{1}{c|}{} & \multicolumn{1}{c|}{\textbf{6.97\%}} & \multicolumn{1}{c|}{\textbf{85.93\%}} &  \\ \hline
\end{tabular}
}
\end{minipage}


\setlength{\tabcolsep}{2pt}
\centering
\resizebox{1.8\columnwidth}{!}{%
\begin{tabular}{|c||c|c|c|c||c|c|c|c||c|c|c|c|}
\hline
 \multicolumn{10}{|r}{ \textbf{ARE} (lower is better) } &\multicolumn{3}{r|}{\barexp%
 }\\ \hline
& \multicolumn{4}{c||}{\texttt{Blackscholes}} & \multicolumn{4}{c||}{\texttt{FFT}} & \multicolumn{4}{c|}{\texttt{Inversek2j}}\\
\hline
{\textbf{Non-cummutative}} 
& \multirow{2}{*}{\textbf{NoSwap}} & \multicolumn{2}{c|}{\textbf{SWAPPER}} & \multirow{2}{*}{\textbf{Theor.}} 
& \multicolumn{1}{c|}{\multirow{2}{*}{\textbf{NoSwap}}} & \multicolumn{2}{c|}{\textbf{SWAPPER}} & \multirow{2}{*}{\textbf{{Theor.}}} 
& \multicolumn{1}{c|}{\multirow{2}{*}{\textbf{NoSwap}}} & \multicolumn{2}{c|}{\textbf{SWAPPER}} & \multirow{2}{*}{\textbf{Theor.}} 
\\ \cline{3-4} \cline{7-8}
\textbf{AxIC multipliers} & & \textbf{Comp.} & \textbf{App.} &  & &
 \multicolumn{1}{c|}{\textbf{Comp.}} & \multicolumn{1}{c|}{\textbf{App.}} &  
 & \multicolumn{1}{c|}{} & \multicolumn{1}{c|}{\textbf{Comp.}} & \multicolumn{1}{c|}{\textbf{App.}} &
 \\ \hline
GQU & \cellcolor{orange!98.2121!red!40}50.89\% & \cellcolor{orange!19.3712!red!40}90.31\% & \cellcolor{green!88.5478!orange!40}5.73\% & \cellcolor{green!88.4844!orange!40}5.76\% 
& \multicolumn{1}{c|}{\cellcolor{orange!90.9765!red!40}54.51\%} & \multicolumn{1}{c|}{\cellcolor{green!86.2618!orange!40}6.87\%} & \multicolumn{1}{c|}{\cellcolor{green!86.2617!orange!40}6.87\%} & \cellcolor{green!86.2618!orange!40}6.87\% 
& \multicolumn{1}{c|}{\cellcolor{green!56.2389!orange!40}21.88\%} & \multicolumn{1}{c|}{\cellcolor{green!75.9469!orange!40}12.03\%} & \multicolumn{1}{c|}{\cellcolor{green!96.1752!orange!40}1.91\%} & \cellcolor{green!96.1566!orange!40}1.92\%
\\ \hline
GQV & \cellcolor{orange!98.1778!red!40}50.91\% & \cellcolor{orange!20.7225!red!40}89.64\% & \cellcolor{green!88.4437!orange!40}5.78\% & \cellcolor{green!88.2302!orange!40}5.88\% 
& \multicolumn{1}{c|}{\cellcolor{orange!90.9552!red!40}54.52\%} & \multicolumn{1}{c|}{\cellcolor{green!86.2518!orange!40}6.87\%} & \multicolumn{1}{c|}{\cellcolor{green!86.2520!orange!40}6.87\%} & \cellcolor{green!86.2518!orange!40}6.87\%
& \multicolumn{1}{c|}{\cellcolor{green!56.2400!orange!40}21.88\%} & \multicolumn{1}{c|}{\cellcolor{green!75.8503!orange!40}12.07\%} & \multicolumn{1}{c|}{\cellcolor{green!95.9801!orange!40}2.01\%} & \cellcolor{green!95.9643!orange!40}2.02\%
\\ \hline
GRU & \cellcolor{orange!93.7487!red!40}53.13\% & \cellcolor{orange!13.5464!red!40}93.23\% & \cellcolor{green!84.4526!orange!40}7.77\% & \cellcolor{green!84.0106!orange!40}7.99\% 
& \multicolumn{1}{c|}{\cellcolor{orange!41.4267!red!40}79.29\%} & \multicolumn{1}{c|}{\cellcolor{green!85.2789!orange!40}7.36\%} & \multicolumn{1}{c|}{\cellcolor{green!85.2786!orange!40}7.36\%} & \cellcolor{green!85.2789!orange!40}7.36\% 
& \multicolumn{1}{c|}{\cellcolor{green!54.9141!orange!40}22.54\%} & \multicolumn{1}{c|}{\cellcolor{green!71.9432!orange!40}14.03\%} & \multicolumn{1}{c|}{\cellcolor{green!91.1677!orange!40}4.42\%} & \cellcolor{green!91.1465!orange!40}4.43\% 
\\ \hline
GSM & \cellcolor{orange!95.1181!red!40}52.44\% & \cellcolor{orange!8.3474!red!40}95.83\% & \cellcolor{green!87.9063!orange!40}6.05\% & \cellcolor{green!88.0537!orange!40}5.97\% 
& \multicolumn{1}{c|}{\cellcolor{orange!10.6006!red!40}94.70\%} & \multicolumn{1}{c|}{\cellcolor{green!86.1689!orange!40}6.92\%} & \multicolumn{1}{c|}{\cellcolor{green!86.1689!orange!40}6.92\%} & \cellcolor{green!86.1689!orange!40}6.92\% 
& \multicolumn{1}{c|}{\cellcolor{green!51.3552!orange!40}24.32\%} & \multicolumn{1}{c|}{\cellcolor{green!76.3324!orange!40}11.83\%} & \multicolumn{1}{c|}{\cellcolor{green!96.8950!orange!40}1.55\%} & \cellcolor{green!96.7772!orange!40}1.61\%
\\ \hline
GV3 & \cellcolor{orange!89.1735!red!40}55.41\% & \cellcolor{orange!10.7891!red!40}94.61\% & \cellcolor{green!80.3234!orange!40}9.84\% & \cellcolor{green!81.1769!orange!40}9.41\% 
&  \multicolumn{1}{c|}{\cellcolor{orange!4.9604!red!40}97.52\%} & \multicolumn{1}{c|}{\cellcolor{orange!68.1340!red!40}65.93\%} & \multicolumn{1}{c|}{\cellcolor{green!84.2879!orange!40}7.86\%} & \cellcolor{green!84.2879!orange!40}7.86\% 
& \multicolumn{1}{c|}{\cellcolor{green!48.1230!orange!40}25.94\%} & \multicolumn{1}{c|}{\cellcolor{green!76.7906!orange!40}11.60\%} & \multicolumn{1}{c|}{\cellcolor{green!97.0017!orange!40}1.50\%} & \cellcolor{green!96.9789!orange!40}1.51\%
\\ \hline
\textbf{\small Gain w.r.t. NoSwap} &  & \textbf{-76.49\%} & \textbf{86.70\%} &  
& \multicolumn{1}{c|}{} & \multicolumn{1}{c|}{\textbf{78.12\%}} & \multicolumn{1}{c|}{\textbf{90.03\%}} & 
& \multicolumn{1}{c|}{} & \multicolumn{1}{c|}{\textbf{46.84\%}} & \multicolumn{1}{c|}{\textbf{90.06\%}} &
\\ \hline
\end{tabular}
}



\setlength{\tabcolsep}{2pt}
\centering
\resizebox{1.8\columnwidth}{!}{%
\begin{tabular}{|c||c|c|c|c||c|c|c|c||c|c|c|c|}
\hline
 \multicolumn{10}{|r}{ \textbf{Average SSIM} (higher is better) } &\multicolumn{3}{r|}{\barexpinv%
 }\\ \hline
& \multicolumn{4}{c||}{\texttt{Kmeans}} & \multicolumn{4}{c||}{\texttt{Sobel}} & \multicolumn{4}{c|}{\texttt{Jpeg}}\\
\hline
\textbf{Non-cummutative}
& \multirow{2}{*}{\textbf{NoSwap}} & \multicolumn{2}{c|}{\textbf{SWAPPER}} & \multirow{2}{*}{\textbf{Theor.}} 
& \multicolumn{1}{c|}{\multirow{2}{*}{\textbf{NoSwap}}} & \multicolumn{2}{c|}{\textbf{SWAPPER}} & \multirow{2}{*}{\textbf{{Theor.}}} 
& \multicolumn{1}{c|}{\multirow{2}{*}{\textbf{NoSwap}}} & \multicolumn{2}{c|}{\textbf{SWAPPER}} & \multirow{2}{*}{\textbf{Theor.}} 
\\ \cline{3-4} \cline{7-8}
\textbf{AxIC multipliers} & & \textbf{Comp.} & \textbf{App.} &  & &
 \multicolumn{1}{c|}{\textbf{Comp.}} & \multicolumn{1}{c|}{\textbf{App.}} &  
 & \multicolumn{1}{c|}{} & \multicolumn{1}{c|}{\textbf{Comp.}} & \multicolumn{1}{c|}{\textbf{App.}} &
 \\ \hline
GQU & \multicolumn{1}{c|}{\cellcolor{orange!90.1750!red!40}0.4509} & \multicolumn{1}{c|}{\cellcolor{orange!95.0849!red!40}0.4754} & \multicolumn{1}{c|}{\cellcolor{green!97.7790!orange!40}0.9889} & \cellcolor{green!97.7790!orange!40}0.9889
& \multicolumn{1}{c|}{\cellcolor{green!99.4602!orange!40}0.9973} & \multicolumn{1}{c|}{\cellcolor{green!99.2455!orange!40}0.9962} & \multicolumn{1}{c|}{\cellcolor{green!99.2455!orange!40}0.9962} & \cellcolor{green!99.2455!orange!40}0.9962
& \cellcolor{green!98.6513!orange!40}0.9933 & \cellcolor{green!99.9451!orange!40}0.9997 & \cellcolor{green!99.9210!orange!40}0.9996 & \cellcolor{green!99.9470!orange!40}0.9997 \\  \hline
GQV & \multicolumn{1}{c|}{\cellcolor{orange!90.1750!red!40}0.4509} & \multicolumn{1}{c|}{\cellcolor{orange!95.0849!red!40}0.4754} & \multicolumn{1}{c|}{\cellcolor{green!96.0177!orange!40}0.9801} & \cellcolor{green!96.0177!orange!40}0.9801 
& \multicolumn{1}{c|}{\cellcolor{green!98.8184!orange!40}0.9941} & \multicolumn{1}{c|}{\cellcolor{green!98.8933!orange!40}0.9945} & \multicolumn{1}{c|}{\cellcolor{green!98.8713!orange!40}0.9944} & \cellcolor{green!98.8933!orange!40}0.9945
& \cellcolor{green!98.6564!orange!40}0.9933 & \cellcolor{green!99.9451!orange!40}0.9997 & \cellcolor{green!99.9179!orange!40}0.9996 & \cellcolor{green!99.9314!orange!40}0.9997 
\\ \hline
GRU & \multicolumn{1}{c|}{\cellcolor{orange!90.1750!red!40}0.4509} & \multicolumn{1}{c|}{\cellcolor{orange!95.6964!red!40}0.4785} & \multicolumn{1}{c|}{\cellcolor{green!73.7441!orange!40}0.8687} & \cellcolor{green!73.7441!orange!40}0.8687 
& \multicolumn{1}{c|}{\cellcolor{green!83.2410!orange!40}0.9162} & \multicolumn{1}{c|}{\cellcolor{green!83.9514!orange!40}0.9198} & \multicolumn{1}{c|}{\cellcolor{green!83.5666!orange!40}0.9178} & \cellcolor{green!83.9514!orange!40}0.9198 
& \cellcolor{green!96.3053!orange!40}0.9815 & \cellcolor{green!99.0550!orange!40}0.9953 & \cellcolor{green!99.7740!orange!40}0.9989 & \cellcolor{green!99.7994!orange!40}0.9990
\\ \hline
GSM & \multicolumn{1}{c|}{\cellcolor{orange!90.1750!red!40}0.4509} & \multicolumn{1}{c|}{\cellcolor{orange!89.7795!red!40}0.4489} & \multicolumn{1}{c|}{\cellcolor{green!95.8918!orange!40}0.9795} & \cellcolor{green!95.8918!orange!40}0.9795 
& \multicolumn{1}{c|}{\cellcolor{green!99.0967!orange!40}0.9955} & \multicolumn{1}{c|}{\cellcolor{green!97.8747!orange!40}0.9894} & \multicolumn{1}{c|}{\cellcolor{green!98.8448!orange!40}0.9942} & \cellcolor{green!97.8747!orange!40}0.9894
& \cellcolor{green!91.1497!orange!40}0.9557 & \cellcolor{green!97.8408!orange!40}0.9892 & \cellcolor{green!99.6295!orange!40}0.9981 & \cellcolor{green!99.6630!orange!40}0.9983
\\ \hline
GV3 & \multicolumn{1}{c|}{\cellcolor{orange!90.1750!red!40}0.4509} & \multicolumn{1}{c|}{\cellcolor{orange!90.2464!red!40}0.4512} & \multicolumn{1}{c|}{\cellcolor{green!82.8110!orange!40}0.9141} & \cellcolor{green!82.8110!orange!40}0.9141  
&  \multicolumn{1}{c|}{\cellcolor{green!97.4255!orange!40}0.9871} & \multicolumn{1}{c|}{\cellcolor{green!83.7149!orange!40}0.9186} & \multicolumn{1}{c|}{\cellcolor{green!97.4255!orange!40}0.9871} & \cellcolor{green!83.7149!orange!40}0.9186  
& \cellcolor{orange!89.6334!red!40}0.4482 & \cellcolor{green!84.8894!orange!40}0.9244 & \cellcolor{green!96.8800!orange!40}0.9844 & \cellcolor{green!97.0556!orange!40}0.9853
\\ \hline
\textbf{\small Gain w.r.t. NoSwap}  & & \multicolumn{1}{|c|}{\textbf{3.33\%}} & \multicolumn{1}{c|}{\textbf{109.87\%}} & 
& \multicolumn{1}{c|}{} & \multicolumn{1}{c|}{\textbf{-1.45\%}} & \multicolumn{1}{c|}{\textbf{-0.005\%}} & 
& \multicolumn{1}{c|}{} & \multicolumn{1}{c|}{\textbf{ 22.49\% }} & \textbf{ 25.42\% } &
\\ \hline
\end{tabular}
}

\vspace{-10pt}

\end{table*}




\subsection{Application level}

\noindent \textbf{Set-up:} We used the CPU benchmarks and their inputs from the AxBench benchmark suite~\cite{Yaz16} (\texttt{Jpeg}, \texttt{Kmeans}, \texttt{Sobel}, \texttt{Blachscholes}, \texttt{FFT}, \texttt{Inversk2j} and \texttt{Jmeint}). The used ViT is \texttt{ViT\_Small} pre-trained on the Imagenet (vit\_small\_patch16\_224 from timm repository~\cite{hugginfaceVIT}). 
\texttt{Jpeg} is implemented with $16$-bit integer arithmetic, while all the others are implemented using floating-point (FP) arithmetic.  To integrate EvoApprox multipliers in the benchmarks using FP, we first translate them to use $32$-bit FxP arithmetic through the libfixmath library.
In the results, we report the impact on accuracy introduced by FxP. 
As no $32$-bit multiplier is currently available in EvoApproxLib, we emulate 32-bit multiplication by applying a typical modular approach~\cite{Kum17}, i.e.:

\vspace{-\baselineskip}

{\small
\begin{equation}\label{eq:32mul}
\begin{split}
A\cdot B &= (A_H 2^n + A_L)\cdot (B_H 2^n + B_L) \\
          &= \underbrace{A_H\cdot B_H 2^{2n}}_{\text{HI}} + \underbrace{A_H\cdot B_L 2^n + A_L\cdot B_H 2^n}_{\text{MD}} + \underbrace{A_L\cdot B_L}_{\text{LO}}
\end{split}
\end{equation}} where $n$ is half the number of the input bits, i.e., 16 for a 32-bit multiplication.
Hence, we can perform $32$-bit signed approximate multiplications using $16$-bit signed EvoAppproxLib multipliers. As the main goal is to evaluate SWAPPER, we use the same approximate multiplier for all 16-bit multiplications.  As we use \textit{mul16s} multipliers (i.e., signed), to comply with Eq.~\ref{eq:32mul} we shift the input values to one position right for $\mathit{MD}$ and $\mathit{LO}$ multiplications.
Depending on which part of the operation we approximate among $\mathit{LO}$, $\mathit{MD}$, and $\mathit{HI}$, we emulate a different approximation of the 32-bit multiplication. Approximating $\mathit{A_H \cdot B_H}$, i.e., $\mathit{HI}$, inserts an absolute error of at least $2^{2n}$, i.e., $2^{32}$ for integer multiplication. From our tests, approximating $\mathit{HI}$ consistently leads to catastrophic errors, making the results unusable according to application-level metrics. Hence, we report results obtained by approximating $\mathit{LO}$ and $\mathit{MD}$.

\begin{figure*}[htp!]
    \centering
\texttt{Jpeg}\par\smallskip
    \begin{subfigure}[t]{0.23\textwidth}
    \centering
    \includegraphics[width=\linewidth]{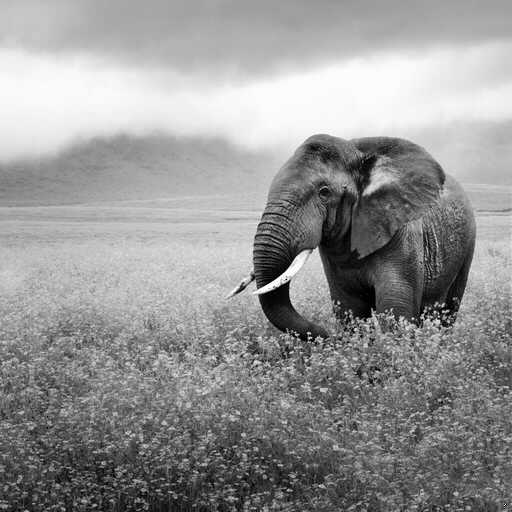}
    \caption{Original \texttt{JPEG} output}
    \label{fig:JPEGgold}
\end{subfigure}
    \hfill
    \begin{subfigure}[t]{0.23\textwidth}
    \centering
    \includegraphics[width=\linewidth]{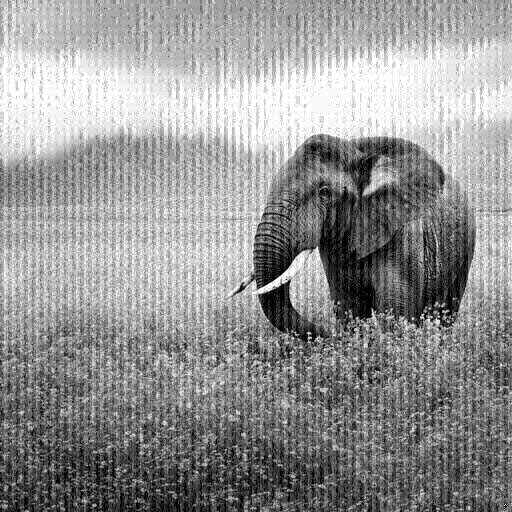}
    \caption{NoSwap (SSIM 0.457)}
    \label{fig:JPEGNoswap}
\end{subfigure}
    \hfill
    \begin{subfigure}[t]{0.23\textwidth}
    \centering
    \includegraphics[width=\linewidth]{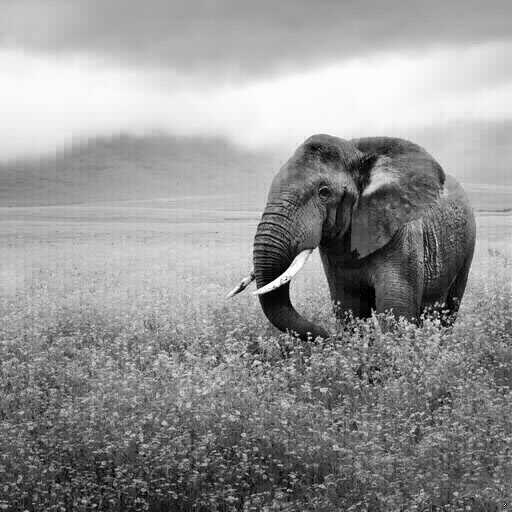}
    \caption{SWAPPER Comp. (SSIM 0.923)}
    \label{fig:JPEGuniform}
\end{subfigure}
    \hfill
    \begin{subfigure}[t]{0.23\textwidth}
    \centering
    \includegraphics[width=\linewidth]{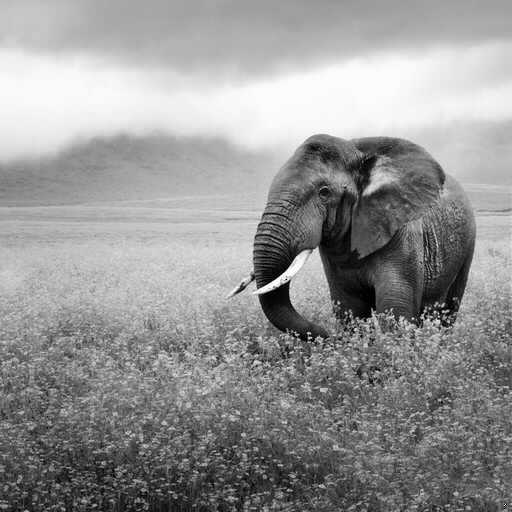}
    \caption{SWAPPER App. (SSIM 0.985)}
    \label{fig:JPEGapptuned}
\end{subfigure}
\par\medskip\texttt{Kmeans}\par\smallskip

\begin{subfigure}[t]{0.23\textwidth}
    \centering
    \includegraphics[width=\linewidth]{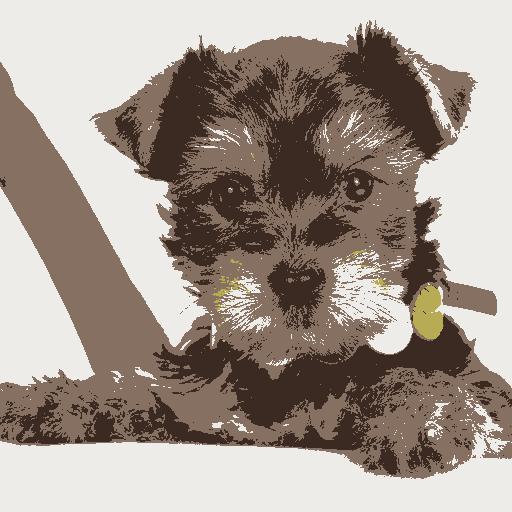}
    \caption{Original \texttt{Kmeans} output}
    \label{fig:kmeansgold}
\end{subfigure}
\hfill
\begin{subfigure}[t]{0.23\textwidth}
    \centering
    \includegraphics[width=\linewidth]{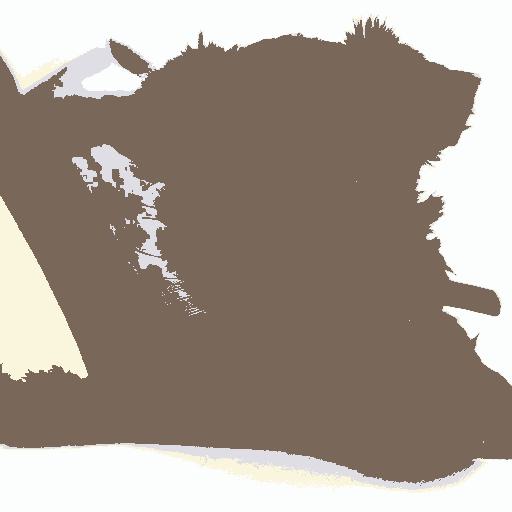}
    \caption{NoSwap (SSIM 0.465)}
    \label{fig:kmeansNoswap}
\end{subfigure}
\hfill
\begin{subfigure}[t]{0.23\textwidth}
    \centering
    \includegraphics[width=\linewidth]{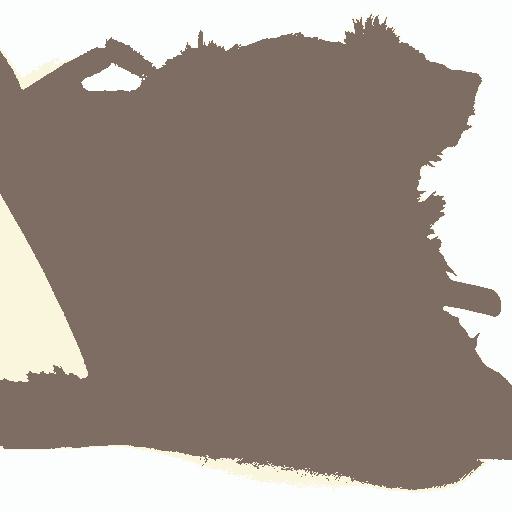}
    \caption{SWAPPER Comp. (SSIM 0.460)}
    \label{fig:kmeansuniform}
\end{subfigure}
\hfill
\begin{subfigure}[t]{0.23\textwidth}
    \centering
    \includegraphics[width=\linewidth]{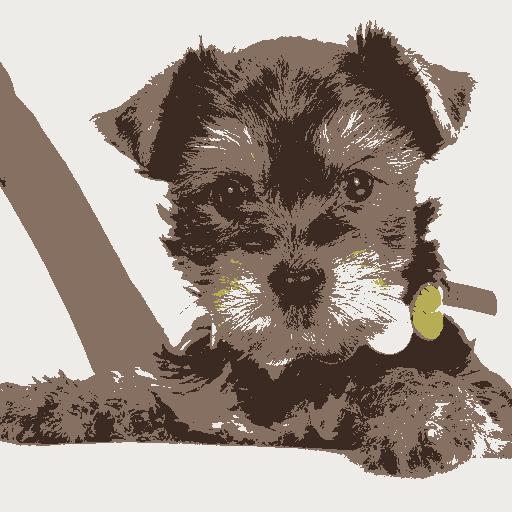}
    \caption{SWAPPER App. (SSIM 0.992)}
    \label{fig:kmeansapptuned}
\end{subfigure}
\par\smallskip
\vspace{-7pt}
\caption{\texttt{Jpeg} and \texttt{Kmeans} output comparison: Original version (\ref{fig:JPEGgold}),(\ref{fig:kmeansgold}); Approximate NoSwap version (\ref{fig:JPEGNoswap}), (\ref{fig:kmeansNoswap});  Approximate SWAPPER version with decision at component level (\ref{fig:JPEGuniform}),(\ref{fig:kmeansuniform}) and application level (\ref{fig:JPEGapptuned}),(\ref{fig:kmeansapptuned}).}
\vspace{-7pt}
\label{fig:all-images}
\end{figure*}

For the application level tuning, we execute the application and measure the error using representative inputs. For benchmarks producing images as outputs, instead of the image difference metric used in~\cite{Yaz16}, we use the more common and advanced Structural Similarity Index Measure (SSIM)\cite{SSIM}. SSIM is a perception-based metric that quantifies similarity between two images by modeling degradation as changes in structural information; perfect similarity gives $SSIM=1$. For the other benchmarks, we rely on AxBench’s metrics: Miss Rate $MR = \frac{\text{Wrong result}}{\text{Total Executions}}$ and Average Relative Error (ARE). For \texttt{ViT}, we use top-1 accuracy. 
As representative inputs, we use the data provided in AxBench
and, for the \texttt{ViT} the ImageNet test set~\cite{hugginfaceVIT}. 
The maximum measured time for the application tuning phase is around 15 minutes for the \texttt{Sobel} benchmark (AMD EPYC 7443, 2.85GHz, without thread parallelization).
For~\texttt{ViT}, using the Transaxx framework~\cite{transaxx_2025}, took approximately 4 hours on average per approximate circuit. For all results, the comparison is done with the Original version of the application.

\noindent {\bf AxBench:} 
The upper left part of Table~\ref{tbl:all-approx-table} reports the results for precise computation of the original (FP) and FxP versions. \texttt{Jpeg} is a special case because it is implemented using 16-bit integer arithmetic. Some benchmarks using the ARE metric exhibit errors even for the original version, due to the way ARE is computed in AxBench
: an error is counted whenever the denominator (i.e., the reference value) equals $0$. We observe some accuracy degradation introduced by the FxP datatype, but without incurring large errors.
For the approximate computation, we report results for 16-bit signed i) commutative multipliers (to which SWAPPER {cannot} be applied) and ii) non-commutative multipliers from EvoApproxLib.  

\noindent \textit{i) Commutative multipliers:} 
The upper left part of Table~\ref{tbl:all-approx-table} reports the error metrics for AxBench benchmarks using commutative multipliers. The induced error is generally high (red/orange), compared to the original version, for all benchmarks except \texttt{Jpeg}.

\noindent \textit{ii) Non-commutative multipliers:}
The rest of able~\ref{tbl:all-approx-table} reports results for the original EvaApproxLib non-commutative multipliers (NoSwap) and for SWAPPER applied at component and application levels (see Section~\ref{sec:proposed}). It also shows SWAPPER’s average gain over NoSwap across all multipliers.
NoSwap yields results comparable to commutative multipliers, except for the \texttt{Blackscholes}, \texttt{Inversek2j}, and \texttt{Sobel} benchmarks. For \texttt{Blackscholes} and \texttt{Inversek2j}, errors are reduced, but they are still far from the original, with an average difference of 53.38\% and 22.25\%, respectively.

With SWAPPER at the component level, several benchmarks nearly match the original version; for example, \texttt{FFT} achieves a 78.12\% average gain and \texttt{Inversek2j} 46.84\% over NoSwap. However, we observe degradation in a few cases, mainly for \texttt{Blackscholes} and slightly for \texttt{Sobel}. In contrast, applying SWAPPER at the application level significantly reduces errors for all benchmarks, yielding results very close to the original version, as indicated by the green cells and high average gains over NoSwap.
SWAPPER with application level tuning nearly matches the theoretical values, with an average difference of only 0.4\%.
``Theor.'' reports application metrics achievable by a component level oracle that always picks the operand order minimizing the multiplier error for each application multiplication, rather than the overall application error. However, even with perfect operand-order selection at the multiplier level, the application may remain highly sensitive to the specific approximation degree of the used multiplier. Thus, these values are not strict upper bounds but should be viewed as indicative of the benefits of operand-order optimization in non-commutative AxIC operators.

Figure~\ref{fig:all-images} presents examples of outputs obtained from different  versions for \texttt{Jpeg} and \texttt{Kmeans} benchmarks. We show two images obtained using the \texttt{16s\_GV3} multiplier for \texttt{Jpeg} and \texttt{16s\_GSM} for \texttt{Kmeans}.
Figures~\ref{fig:JPEGgold} and~\ref{fig:kmeansgold} show the original output. As depicted by Figures~\ref{fig:JPEGNoswap} and~\ref{fig:kmeansNoswap}, the output of the approximate version with the original EvaApproxLib multipliers (NoSwap) is of low quality, having an SSIM value of $0.457$ (\texttt{Jpeg}) and  $0.465$ (\texttt{Kmean}). Applying the proposed SWAPPER at the component level, the output results are improved by {$102\%$} for \texttt{Jpeg}, but not for \texttt{Kmeans}, as shown in Figures~\ref{fig:JPEGuniform} and~\ref{fig:kmeansuniform}. However, when SWAPPER is applied at the application level, the results are significantly improved by $116\%$ for \texttt{Jpeg} and  $113\%$ for \texttt{Kmeans}, as Figures~\ref{fig:JPEGapptuned} and~\ref{fig:kmeansapptuned} show. 
\begin{figure*}[h]
\setlength{\abovecaptionskip}{5pt} 
    \centering
\includegraphics[scale=0.6]{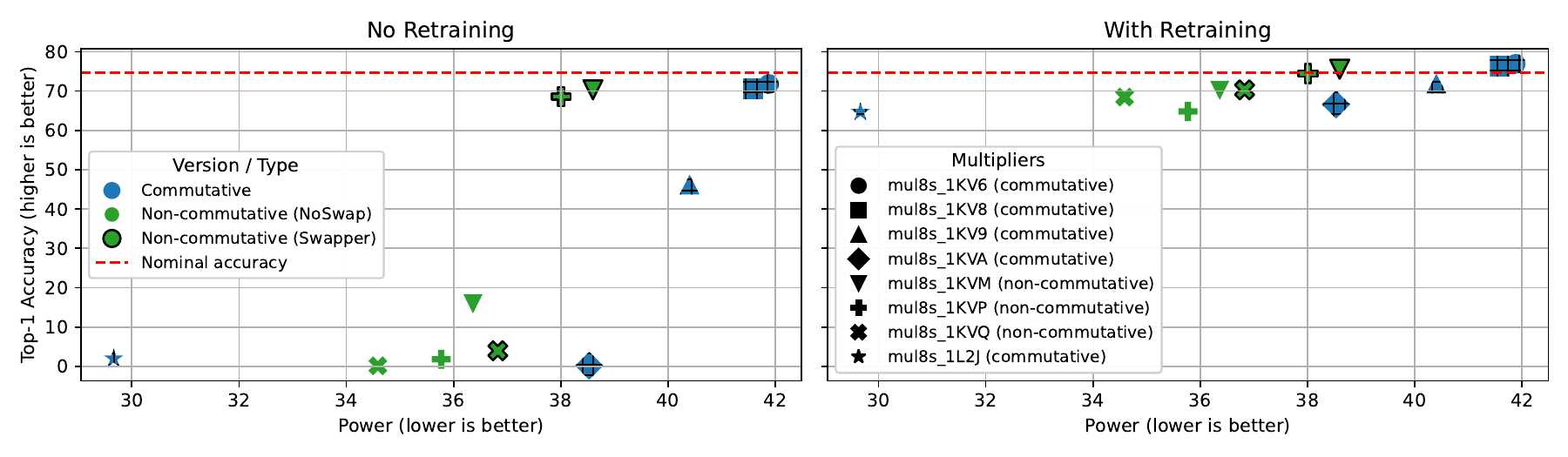}
\vspace{-20pt}
\caption{\texttt{ViT} Top-1 accuracy: Original, commutative and non-cummutative \texttt{8s\_} multipliers from EvoApproxLib~\cite{evoapprox}.}
\label{fig:VIT-results}
\vspace{-10pt}
\end{figure*}

\noindent {\textbf{ViT}:}
As Figure~\ref{fig:VIT-results} shows, the precise \texttt{ViT} version (pre-trained) has 74.64\% top-1 accuracy.
For the approximate versions, we used 8-bit signed multipliers from EvoApproxLib.
For the original EvoApproxLib, only the commutative \texttt{1KV6} and \texttt{1KV8} multipliers managed to reach an accuracy close to the original, and none of the original non-commutative multipliers (\texttt{1KVM}, \texttt{1KVP}, \texttt{1KVQ}). SWAPPER manages to bring the top-1 accuracy close to the original for \texttt{1KVM} and \texttt{1KVP}. 
Furthermore, we applied retraining (maximum 60 epoches) to fine-tune the model with the introduced approximation. SWAPPER still manages to improve the accuracy compared original EvoApproxLib multipliers. Note that, for the \texttt{1L2J}, \texttt{1L2L}, and \texttt{1L2N} commutative multipliers, and the \texttt{1KX5} and \texttt{1KXF} non-commutative ones, retraining, even for 60 epochs, failed to recover the accuracy lost due to extreme approximation introduced by these AIxCs. Table~\ref{tab:mul} shows that these multipliers have high MAE (higher than 50).

\begin{figure*}[h]
    \centering
    \setlength{\abovecaptionskip}{5pt} 
    \includegraphics[width=\textwidth]{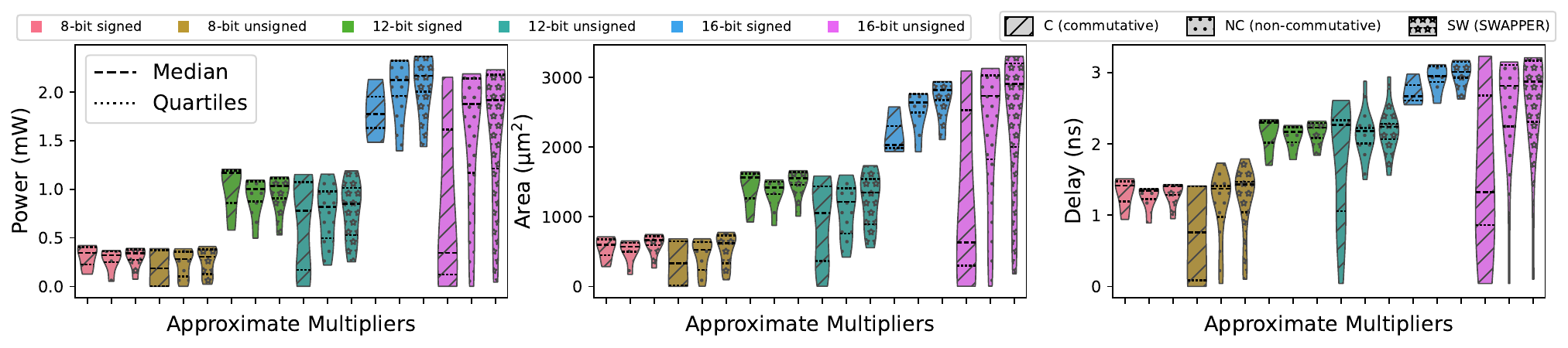}
\vspace{-15pt}
\caption{Power, Area and Delay violin plots for 8-, 12- and 16-bit signed and unsigned NC, C and SWAPPER multipliers}
\label{fig:violin-power-area-delay}
\vspace{-10pt}
\end{figure*}


\begin{table}[h]
\caption{Synthesis results for a possible  HW implementation}
\vspace{-5pt}
\label{tab:swapper-HW}
\setlength{\tabcolsep}{2pt}
\centering
\resizebox{\columnwidth}{!}{%
\begin{tabular}{cccccccccc}
 & \multicolumn{3}{c}{\textbf{8-bit}} & \multicolumn{3}{c}{\textbf{12-bit}} & \multicolumn{3}{c}{\textbf{16-bit}} \\ \cline{2-10} 
\multirow{2}{*}{\textbf{\begin{tabular}[c]{@{}c@{}}HW\\ Swapper\end{tabular}}} & \textbf{P} & \textbf{A} & \multicolumn{1}{c|}{\textbf{D}} & \textbf{P} & \textbf{A} & \multicolumn{1}{c|}{\textbf{D}} & \textbf{P} & \textbf{A} & \textbf{D} \\
 & 0.0227 & 92.9214 & \multicolumn{1}{c|}{0.06} & 0.033 & 134.2198 & \multicolumn{1}{c|}{0.06} & 0.0429 & 175.5182 & 0.06 \\ \cline{2-10} 
\textbf{} & \multicolumn{9}{l}{\textbf{P = Power (mW); A = Area ($\mathbf{\mu m^2}$); D = Delay (ns)}}
\end{tabular}
}
\vspace{-10pt}
\end{table}

\subsection{Hardware Implementation}

In this subsection, we provide a possible hardware implementation of the swapping logic, composed of two multiplexers swapping the two input operands based on a 1-bit signal.
We synthesized it using a 45nm PDK technology library, as in EvoApproxLib~\cite{evoapprox}. 
Table~\ref{tab:swapper-HW} reports the synthesis result of such design for 8-, 12-, and 16-bit operands in terms of power, area, and delay. 

We implemented the swapping logic on EvoApproxLib non-commutative multipliers to assess its overhead. Fig.~\ref{fig:violin-power-area-delay} compares the HW attributes of the complete set of EvoApproxLib with commutative (C), non-commutative (NC), and non-commutative with swapping logic (SW) multipliers. Overall, SWAPPER has limited impact on power, area, and delay. For 8-bit multipliers, the average overhead is about 10\% power, 22\% area, and 5\% delay, while for 16-bit multipliers it drops to about 2\% power, 8\% area, and 2\% delay.

\begin{figure}[tp]
    \centering
    \includegraphics[width=\columnwidth]{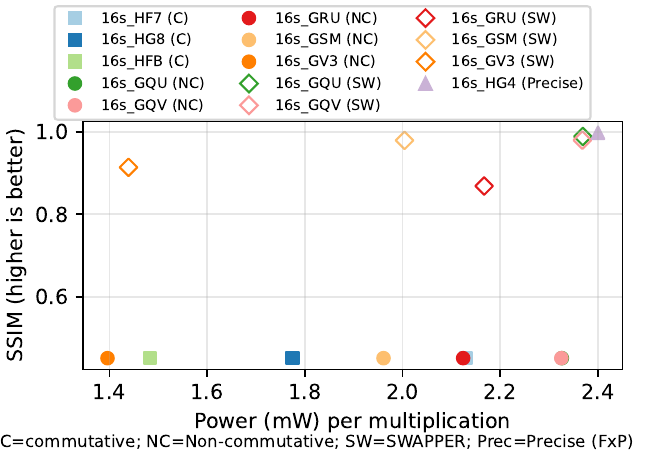}
\vspace{-20pt}
\caption{\texttt{Kmeans}: Power/multiplication vs. SSIM (\texttt{16\_s} mult.). 
}
\label{fig:kmeans-power-ssim}
\vspace{-10pt}
\end{figure}

Overall, SWAPPER is able to improve application accuracy compared to the original EvoApproxLib multipliers with a little power overhead. We used TransAxx framework~\cite{transaxx_2025}
to compare the error introduced due to approximation and the Power per multiplication, for all multipliers. All values regarding the specifications of each multiplier were obtained by EvoApproxLib and our SWAPPER synthesis result. 
As an example, Fig.~\ref{fig:kmeans-power-ssim} shows the trade-off between Power per multiplication and SSIM for the \texttt{Kmeans} benchmark using commutative (\drawsquare{000000}) and non-commutative (\drawdot{000000}) 16-bit signed approximate multipliers, compared to SWAPPER (\drawdiamond{000000}). The baseline results are with a 16-bit signed precise multiplier (\precise{CAB2D6}). Both commutative and non-commutative show similar results in terms of SSIM (average 0.45). However, by applying SWAPPER with application level tuning, it is possible to obtain significant increases in the SSIM -- relative gain ranging between $92.66\%$ and $119.31\%$ -- compared to not applying it. For instance, when SWAPPER is applied to \texttt{mul16s\_GSM} (see \mul{FDBF70} and \drawdiamond{FDBF70}), it increases the SSIM from 0.4509 \mul{FDBF70} to 0.9797 \drawdiamond{FDBF70} (117.2\% relative gain) with a slight power increase, from 1.961 \mul{FDBF70} to 2.004 mW \drawdiamond{FDBF70} ($\approx2\%$). The obtained SSIM is very close to the precise FxP version (i.e., 0.998, see \precise{CAB2D6}), with considerably less consumption than the precise multiplier (2.4 mW \precise{CAB2D6}).

\section{Conclusions and future directions}

We exploit the varying error profile property of non-commutative approximate circuits to introduce SWAPPER, a lightweight approach that dynamically swaps operands using a single-bit decision mechanism, significantly reducing approximation error.
An exploration framework identifies the optimal decision bit, achieving significant accuracy improvements at both component and application levels. Future work will explore heterogeneous approximate operators and additional arithmetic properties to further reduce online errors.

\vspace{-10pt}
\begin{acks}
Supported by the French National Research Agency (ANR) -- REAxION (ANR-25-CE25-5926) and RADYAL (ANR-23-IAS3-0002) projects.
\end{acks}

\bibliographystyle{ACM-Reference-Format}
\bibliography{refs.bib}

\end{document}